\documentclass[aps,prl,twocolumn,showpacs,superscriptaddress]{revtex4}
\usepackage{amsmath}
\usepackage{dcolumn}
\usepackage{epsfig}
\usepackage{graphicx}
\usepackage{latexsym}
\usepackage{amssymb}

\begin{document}
\title{The approach to a superconductor-to-Bose-insulator transition in disordered films}

\author{Myles A. Steiner}
\altaffiliation{Present address: National Renewable Energy Laboratory, Golden, CO 80401}
\affiliation{Department of Applied Physics, Stanford University, Stanford, CA 94305}
\author{Nicholas P. Breznay}
\affiliation{Department of Applied Physics, Stanford University, Stanford, CA 94305}
\author{ Aharon Kapitulnik} 
\affiliation{Department of Applied Physics, Stanford University, Stanford, CA 94305} 
\affiliation{Department of Physics, Stanford University, Stanford, CA 94305}

\date{\today}

\begin{abstract} 
Through a detailed study of scaling near the magnetic field-tuned superconductor-to-insulator transition in strongly disordered films, we find that results for a variety of materials can be collapsed onto a single phase diagram. The data display two clear branches, one with weak disorder and an intervening metallic phase, the other with strong disorder. Along the strongly disordered branch, the resistance at the critical point approaches $R_Q = h/4e^2$ and the scaling of the resistance is consistent with quantum percolation, and therefore with the predictions of the dirty boson model.
\end{abstract}

\pacs{74.20.M, 74.76.-W, 73.40.Hm }

\maketitle

At finite magnetic field, a disordered superconducting system, thinned down to a two-dimensional film, may exhibit a true superconducting state only at zero temperature.  Upon increasing either the magnetic field or the disorder superconductivity will be lost \cite{ffh1991}.  Two main scenarios have been proposed for this transition. In the ``dirty boson" scenario Cooper-pairs exhibit enough integrity for the transition to be dominated by boson localization \cite{mpaf1990}. Alternatively, pairs are broken at the transition and the system above the transition is dominated by fermion physics \cite{finkelshtein1987}. In either scenario it is expected that above the transition a large enough system will exhibit insulating behavior as the temperature approaches zero. Focusing on the magnetic field-tuned transition, initial experiments seemed to confirm the boson-dominated scenario \cite{goldman1998}. However, scaling analysis around the apparent superconductor-insulator transition (SIT) \cite{yazdani1995,mason1999,steiner2005phc} yielded critical exponents that resemble those of classical rather than quantum percolation, and the critical resistance at the transition was not the quantum of resistance for Cooper pairs, $R_Q \equiv h/4e^2 \approx 6.45$ k$\Omega/\Box$ \cite{fgg1990}. In addition, some experiments have challenged the general occurrence of a SIT, demonstrating an apparent transition that is only a crossover to a new metallic state at low temperatures. In all these cases the insulating phase above the transition was rather weak, suggesting the possibility that nearby fermions couple dissipatively to the system, hence preventing the observation of a pure boson-dominated SIT \cite{kapitulnik2001}. More recent experiments on amorphous indium-oxide (InOx) \cite{steiner2005phc,samb2004} and TiN fims \cite{baturina} indicated that the strength of the insulating behavior above the apparent magnetic field-tuned SIT can be enhanced by using more disordered films \cite{steiner2005phc}. However, no systematic study of the critical behavior was undertaken to shed light on the quest to achieve a true boson-dominated SIT.

\begin{figure}[h]
\centering
\includegraphics[width=0.95\columnwidth]{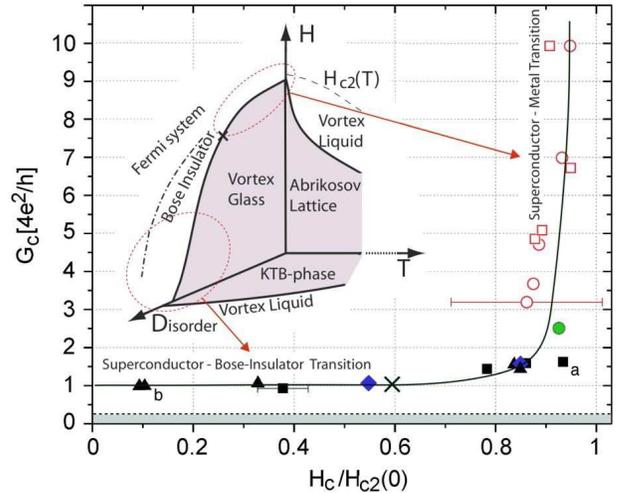}
\caption{ \footnotesize \setlength{\baselineskip}{0.8\baselineskip} Phase diagram of all samples including InOx and MoGe using reduced axes. Open circles (Ref.~\cite{masonthesis2001}) and squares (Ref.~\cite{yazdanithesis1994}) are for MoGe films. Full circle is for thin Ta films (Ref.~\cite{yoonta}). Diamonds are for InOx (Refs.~\cite{hp1990,hpr1992}) and full triangles and squares are new InOx data for this work, from two batches. The error bars give a measure of the difficulty in determining $H_{c2}(0)$. The solid line is a guide to the eye. The inset shows the low temperature part of the generic phase diagram as proposed in Ref.~\cite{mpaf1990}. $\times$ denotes a possible critical point between a transition to a Bose insulator and a transition to a metal (see text). The bottom shaded area  denotes the region for which the conductance is smaller than $h/e^2$. Samples ``a" and ``b" are discussed further in Fig.~\ref{samples}. Different regimes of the diagram as related to the data are explained in the text.}
\label{all}
\end{figure}

In this paper we study the SIT critical behavior of a wide range of InOx films at temperatures well below their mean-field superconducting transition (denoted as $T_{c0}$). While different batches of samples may have different properties for similar normal-state $R_{\Box}$ \cite{samples}, we found that all samples can be collapsed onto a single phase diagram if we plot the critical resistance at the SIT normalized by $R_Q$ versus the critical field at the transition normalized by $H_{c2}(0)$ (i.e. the mean-field upper critical field at $T=0$). Moreover, on the same phase diagram other samples from previously reported measurements on amorphous MoGe can be placed as well.  Based on our observations we conclude that a true ``dirty-boson" SIT can be achieved in the limit of strong disorder. In that limit the critical magnetic field approaches zero while the critical resistance approaches the quantum of resistance for pairs. Scaling near the critical point, as viewed from measurements at finite temperatures, is consistent with a quantum percolation critical point.  In the opposite limit of weak disorder, such as our earlier measurements on MoGe \cite{yazdani1995,mason1999}, the critical field approaches $H_{c2}(0)$ and the critical resistance decreases with respect to $R_Q$ until it is indistinguishable from the normal-state resistance of the sample. In that limit scaling produces classical percolation critical behavior that is disrupted by the appearance of an intervening metallic phase. Figure~\ref{all} is a summary of our result. It provides a new perspective on the SIT quantum critical behavior, suggesting that while a true Bose-like SIT can be achieved in the limit of strong disorder, the other limit represented by the asymptote $H_c/H_{c2}(0) =1$ contains completely new physics for which a new quantum metallic state could be the featured highlight \cite{oreto}.

InOx films reported in this paper were fabricated by electron beam evaporation of a $99.999 \%$ pure indium oxide piece onto cleaned silicon nitride substrates \cite{steinerthesis} . The disorder was tuned during fabrication by controlling the sample thickness and the oxygen pressure \cite{steiner2005phc}. The films were subsequently patterned into four-point resistivity bars by photolithography and argon-ion etching. Care was taken not to heat the samples above $\sim 50^0$C in order to prevent any recrystallization \cite{kowal1994}. Resistance measurements were conducted in a dilution refrigerator using standard four-point lockin techniques.  Over 15 samples were used for the present study.  The mean-field transition temperature $T_{c0}$, an energy scale for the formation of Cooper pairs in these strongly fluctuating systems, was determined by fitting the fluctuation conductivity to the Aslamasov-Larkin model for two dimensions \cite{aslark1968}. Within each batch of samples disorder increases as $T_{c0}$ decreases and the normal resistance increases. The upper critical field was determined from the low field initial slope of $T_c(H)$ as explained in detail in \cite{steiner2005phc}.

Upon increasing the magnetic field, the behavior of the films changes from superconducting-like with resistance decreasing with decreasing temperature, to insulating-like with resistance increasing with decreasing temperature. The separating curve marks the SIT, defining a critical resistance $R_c$ and magnetic field $H_c$.  Isotherms of resistance versus magnetic field around this point all cross at $H_c$  with a critical resistance $R_c$ marking a zero-temperature quantum SIT phase transition.  Assuming a magnetic field-tuned transition, the correlation length diverges at the transition as  $\xi(H) \propto |H - H_c|^{-\nu}$. The dynamics of the system are characterized by the relation between the correlation length and a vanishing frequency $\Omega$ through $\Omega \propto \xi^{-z}$.  At a finite temperature, $\hbar \Omega$ is cut-off by $k_BT$, defining a length $L_T$ for the cut-off of quantum fluctuations: $k_BT \sim L_T^{-z}$. This leads to a scaling relation for the resistance which is  commonly written as \cite{mpaf1990}

\begin{equation}
R(H,T) = R_c \mathcal{F}_T\left((H-H_c)/T^{1/z\nu}\right).
\end{equation}

\begin{figure}[ht]
\centering
\includegraphics[width=0.9\columnwidth]{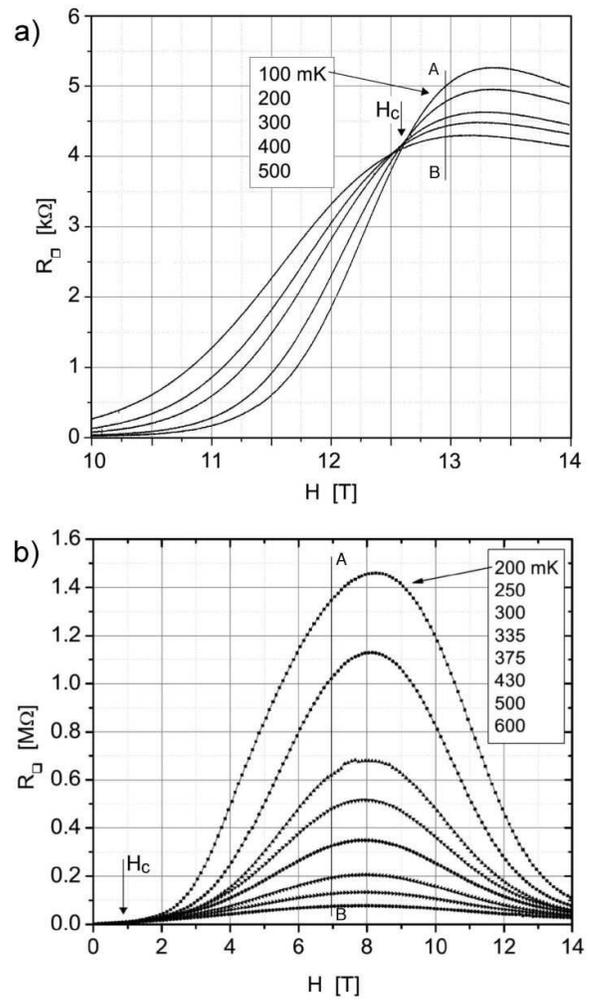}
\caption{ \footnotesize \setlength{\baselineskip}{0.8\baselineskip}  Resistance isotherms for the two samples marked in Fig.~\ref{all}. The isotherm temperatures are listed in the legend. Note the different scales of the vertical axis. Arrow points to the critical fields of $H_c = 12.6$ T for sample ``a", and  $H_c = 0.83$ T for sample ``b".  Line A-B denotes the field at which the activation energy (as determined from the resistance vs. temperature at a fixed field) is maximum.}
\label{samples}
\end{figure}

Figure~\ref{samples} shows the resistance isotherms in a perpendicular magnetic field for two InOx samples, one with weak disorder (Fig.~\ref{samples}a) and one with very strong disorder (Fig.~\ref{samples}b). Similar types of samples were introduced earlier \cite{steiner2005phc}.  For the sample in Fig.~\ref{samples}a the critical resistance is $\sim$4.1 k$\Omega$, very close to the normal state resistance before superconductivity sets in, and the magnetoresistance peak is weak.  Furthermore, the critical field is $\sim$12.6 T, very close to $H_{c2}(0)\approx$ 14 T calculated for this sample. Thus, this sample is very similar to the MoGe samples  previously reported \cite{yazdani1995,mason1999}. Scaling of this sample, as well as all previously reported MoGe samples, typically fails at the lowest temperatures. However, in a finite temperature range scaling yields exponents similar to classical percolation (in particular $\nu \approx 1.35$). 

\begin{figure}[ht]
\centering
\includegraphics[width=0.9\columnwidth]{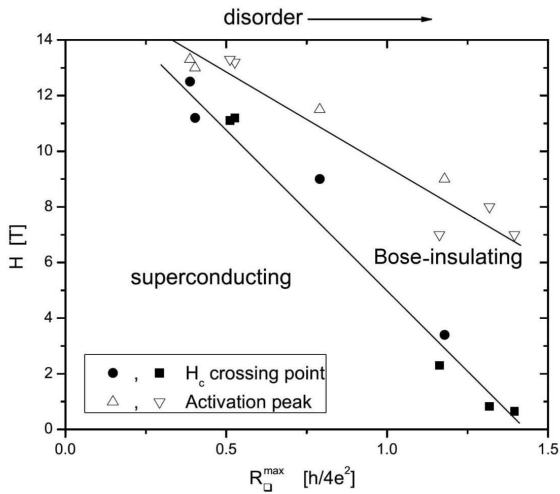}
\caption{ \footnotesize \setlength{\baselineskip}{0.8\baselineskip}  Phase diagram for the superconducting and insulating phases. The various closed and open symbols denote the $H_c$ crossing and the $H_p$ activation peak, respectively, accounting for the two batches of samples. The lines are guides to the eye.}
\label{comp}
\end{figure}

Starting from the crossing point and increasing the magnetic field towards the magnetoresistance peak we note in Fig.~\ref{samples}  that the resistance increases with decreasing temperature, at an increasing rate. At some field this rate is maximum (denoted by lines (A-B)) and we can calculate an activation energy through this magnetic field cut ($\equiv H_p$, also see \cite{steiner2005phc}). To evaluate the trend of the samples we also need a measure of the disorder. For each sample we take the maximum normal-state resistance at zero magnetic field, before superconductivity sets in, as such measure, denoted $R^{max}_\Box$. In Fig.~\ref{comp} we plot $H_c$ and $H_p$ versus $R^{max}_\Box$, and we clearly identify three regimes: below the crossing point the phase is superconducting with resistance decreasing with decreasing temperature; at very high fields, above the field of maximum activation energy, the system is expected to be dominated by Fermi physics; in between the two sets of points the system behaves as a Bose insulator \cite{mpaf1990}. The two solid lines, though drawn solely as guides to the eye, are not parallel but get farther apart as the disorder increases, indicating that the Bose-insulating phase is becoming more well-developed and less influenced by background fermions \cite{yazdani1995}. This is in contrast to figure 5 in Ref.~\cite{samb2004} which uses a different measure for the sample disorder and is drawn over a smaller range of critical fields. 

\begin{figure}[ht]
\centering
\includegraphics[width=0.9\columnwidth]{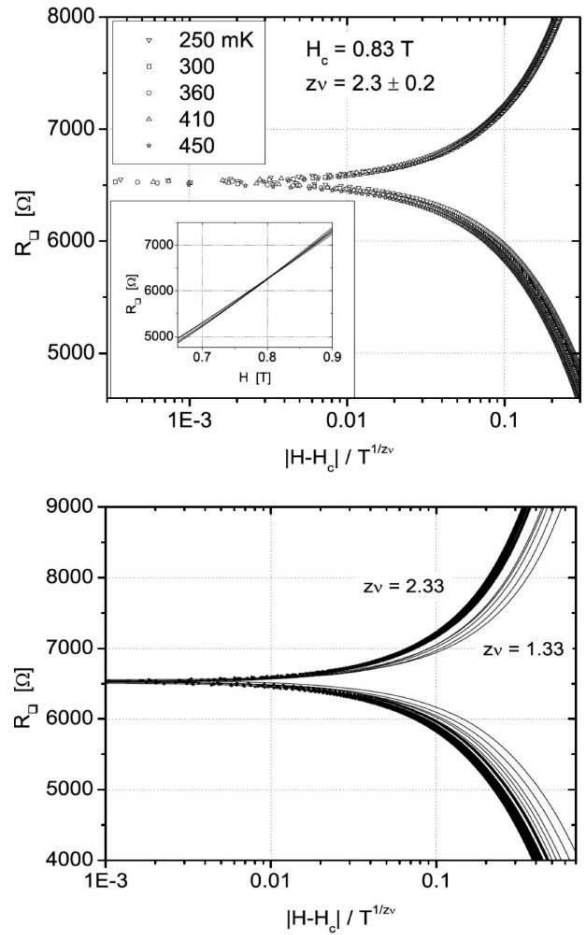}
\caption{ \footnotesize \setlength{\baselineskip}{0.8\baselineskip}  Top: Temperature scaling for sample ``b" above. $R_c$ and $H_c$ are determined from the isotherms' crossing point. The resulting exponent $z\nu$ is determined from the fit.  The inset shows the crossing of the isotherms.  Bottom: The same data as above but with scaling enforced with theoretical models for classical percolation ($z\nu =4/3$) and quantum percolation ($z\nu =7/3$).}
\label{scale}
\end{figure}

InOx films appear to complement MoGe films in terms of the strength of the disorder, the two systems together spanning a wide range of the disorder axis. We now consider InOx samples for which the critical field $H_c$ is very low and the regime of Bose-insulating behavior is very large, and we analyze the critical behavior near the SIT. We have attempted to connect the results with materials of lower disorder such as MoGe, thin Ta films \cite{yoonta}, and other InOx samples such as in reference \cite{hp1990}. While many systems have shown scaling near the SIT with $z\nu \approx 1.3$ \cite{hp1990,yazdani1995,mason1999,steiner2005phc}, no scaling consistent with what is expected for a true ``dirty boson" system has been demonstrated. Figure~\ref{scale} shows the scaling obtained for the highly disordered sample of Fig.~\ref{samples}b. We note that this sample exhibits very low $H_c/H_{c2}(0)$ and a critical resistance of $R_c \approx 6.5$ k$\Omega$. Scaling near the crossing point was obtained in two steps. First, a best fit in which the exponent product $z\nu$ was a free variable was employed. The scaling for each data set was based on its own actual crossing point. Following the procedure described previously \cite{hp1990,yazdani1995,mason1999,steiner2005phc}, the exponent product for this sample was found to be $z\nu = 2.3 \pm 0.2$ (see Fig.~\ref{scale}a). We note that this product is what we normally get in scaling for the low resistance films, plus one, pointing towards a quantum percolation exponent. However, to test this result against specific theoretical predictions, we analyzed the data with two fixed values for $z\nu$ predicted by percolation theory: $z\nu=4/3$ for classical percolation and $z\nu=7/3$ for quantum percolation.  It is evident from Fig.~\ref{scale}b that the higher exponent allows for significantly less spread of the data, reinforcing the free-fit result of 2.3 rather than a lower exponent.  We have carried out a similar analysis on several other films of varying degrees of disorder \cite{steinerthesis}. For other highly disordered films we found the scaling exponent to be consistent with $\nu = 7/3$. For the significantly more weakly disordered samples (see \cite{steiner2005phc}, or the sample in Fig.~\ref{samples}a) the scaling exponent was found to be $\nu \approx 1.3$ with a sometimes broadened crossing point and poorer scaling quality. Thus, the results in this regime essentially mimic those of MoGe.  We have assembled all of our data, as well as various data from the literature, into the plot shown in Fig.~ \ref{all}.

Let us finally return to a detailed discussion of Fig.~\ref{all}. The vertical axis gives the conductance at the critical point in units of $4e^2/h$. The horizontal axis gives the field at the critical point, normalized by the estimated $H_{c2}(0)$; the disorder decreases in the positive direction. For the InOx samples, $H_{c2}(0)$ is calculated from the slope near $T_{c0}$ with some localization corrections as discussed in  \cite{steiner2005phc}; this field was also shown to be close to the peak of the magnetoresistance at low temperatures. For the MoGe samples $H_{c2}$ was measured from the low-field superconducting transitions and the values are listed in references \cite{yazdanithesis1994} and \cite{masonthesis2001}. For the tantalum films, a discussion of the critical field and the coherence length is given in reference \cite{yoonta}. The error bars in Fig.~\ref{all} indicate the difficulty in determining $H_{c2}$ in the weakly insulating films. 

The data clearly display two branches. The samples in the top-right branch are very weak insulators with a critical conductance found to be well above the predicted [6.45 k$\Omega$]$^{-1}$ and a critical field in close proximity to the upper critical field $H_{c2}$. The scaling for all of these samples was found to be consistent with the classical percolation model, with $z = 1$ and $\nu \approx 1.35$. The samples in the lower-left branch, on the other hand, are increasingly strong insulators with a critical conductance approximately the universal value, and scaling clearly shown in this paper to be consistent with $\nu \approx 7/3$. The samples near where the two branches meet (low $G_c$ but large $H_c/H_{c2}$) show scaling that is more consistent with $\nu = 4/3$ than with 7/3 but of generally poor quality, which likely represents some crossover regime between the two branches. We therefore assert that there is a point ($G_c^*,H_c^*$) on our phase diagram which separates the tendencies towards insulating and metallic behavior, and we represent this point in Fig.~\ref{all} by $\times$.

In summary, as the disorder increases, the predictions of the dirty boson model and the expected quantum critical behavior are clearly observed. Given the overall trend that the insulating strength increases as the disorder increases, thereby suppressing any pair-breaking in the vicinity of the crossing point, it would appear that the assumption behind the dirty boson model \textemdash that the carriers are exclusively charge-$2e$ bosons \textemdash is increasingly valid. We expect that in the limit of the most strongly disordered films, the predicted quantum critical behaviour should be observable down to lowest temperature. At the same time, the fact that the low disorder data clumps along a single branch maybe an indication of new physics in that regime. Such new physics \cite{oreto} may be able to explain the widely observed metallic phase for high conductance films.

\noindent {\bf Acknowledgments:}   It is a pleasure to thank Steven Kivelson for very stimulating discussions. Work supported by the National Science Foundation Grant:  NSF-DMR-9508419 and by a NSF Graduate Research Fellowship (NPB).


\begin{references}
\bibitem{samples}
While the relevant axis is disorder, this is difficult to quantify because it is not simply the microscopic disorder that determines R$_\Box$, but a more complicated one depending, for example, on the pinning potential. Thus ``different batches" means that samples may have different properties for similar normal-state R$_\Box$.
\end{references}
\end{document}